\documentclass[manuscript,screen]{acmart}

\AtBeginDocument{%
  }

\setcopyright{acmlicensed}
\copyrightyear{2018}
\acmYear{2018}
\acmDOI{}

\acmISBN{978-1-4503-XXXX-X/18/06}

\usepackage{stmaryrd}
\usepackage{listings}
\usepackage{algorithm}
\usepackage{algpseudocode}
\usepackage{multirow}
\usepackage{color}
\usepackage{multicol}
\usepackage{wrapfig}
\usepackage{graphicx}
\usepackage[lofdepth,lotdepth]{subfig}
\usepackage{time}
\usepackage{soul}
\usepackage{cancel}
\usepackage{prftree}

\newcommand{\oomit}[1]{}
\newcommand\ode[2]{\langle #1 \mathop{\&} #2\rangle}

\newcommand\inv[3]{\llbracket #1 \rrbracket \langle #2 \rangle \llbracket #3\rrbracket}

\begin{document}

\title{Mars 2.0: A Toolchain for Modeling, Analysis, Verification and Code Generation of Cyber-Physical Systems}

 \author{Bohua Zhan}
\email{bzhan@ios.ac.cn}
\author{Xiong Xu}
\email{xux@ios.ac.cn}
\author{Qiang Gao}
\email{gaoqiang@ios.ac.cn}
\author{Zekun Ji}
\email{jizk@ios.ac.cn}
\author{Xiangyu Jin}
\email{jinxy@ios.ac.cn}
\author{Shuling Wang}
\authornote{Corresponding author}
\email{wangsl@ios.ac.cn}
\author{Naijun Zhan}
\email{znj@ios.ac.cn}
\affiliation{%
  \institution{State Key Lab. of Computer Science, Institute of Software, Chinese Academy of Sciences}
  \streetaddress{Zhongguancun South 4th Street, No 4, Haidian District}
  \city{Beijing}
  \country{China}
  \postcode{100190}
}

\renewcommand{\shortauthors}{Zhan et al.}

\lstdefinelanguage{hcsp}{
	language=python,                
	basicstyle=\ttfamily\scriptsize,
	numbers=left,                   
	numberstyle=\tiny,              
	numbersep=5pt,                  
	showtabs=false,                 
	tabsize=2,                      
	captionpos=b,                   
	breaklines=true,                
	breakatwhitespace=true,         
	escapeinside={(*}{*)},        
	extendedchars=false,
	keywordstyle=\textcolor[rgb]{0.5,0,0.35},
	stringstyle=\textcolor[rgb]{0.6,0,0},
	commentstyle=\textcolor[rgb]{0.25,0.5,0.35},
	belowskip=-3em,
	morekeywords={end,if,elif,end,then,endif, while, in, for,<,>,>=,<=,==,-->,put,get_highest,delete,push,pop,head,tail,do,endwhile, module, begin, endmodule, procedure}
}

\lstdefinelanguage{aadl}{
	language=java,                
	basicstyle=\ttfamily\footnotesize,       
	numbers=left,                   
	numberstyle=\tiny,              
	numbersep=5pt,                  
	showspaces=false,               
	showstringspaces=false,         
	showtabs=false,                 
	tabsize=2,                      
	captionpos=b,                   
	breaklines=true,                
	breakatwhitespace=true,         
	escapeinside={(*}{*)},        
	extendedchars=false,
	keywordstyle=\textcolor[rgb]{0.5,0,0.35},
	stringstyle=\textcolor[rgb]{0.6,0,0},
	commentstyle=\textcolor[rgb]{0.25,0.5,0.35},
	belowskip=-3em,
	morekeywords={package, system,end,implementation,if,elsif,end,processor,memory,bus,device,thread,process,annex,abstract,features,in,out,data,port,event,properties,invariant,variables,states,transition,connections,subcomponents,initial,complete,final,and,assert,<,>}
}

\lstdefinelanguage{transC}{
	language=java,                
	basicstyle=\ttfamily\footnotesize,       
	numbers=left,                   
	numberstyle=\tiny,              
	numbersep=5pt,                  
	showspaces=false,               
	showstringspaces=false,         
	showtabs=false,                 
	tabsize=2,                      
	captionpos=b,                   
	breaklines=true,                
	breakatwhitespace=true,         
	escapeinside={(*}{*)},        
	extendedchars=false,
	keywordstyle=\textcolor[rgb]{0,0,1.0},
	stringstyle=\textcolor[rgb]{0.6,0,0},
	commentstyle=\textcolor[rgb]{0.25,0.5,0.35},
	belowskip=-3em,
	morekeywords={typedef,struct, if,elsif,end,processor,memory,bus,device,thread,process,annex,abstract,features,}
}

\newcommand{\aadlss}{\textsf{AADL}\!\oplus\!\textsf{S/S}}

\newcommand{\pskip}{\textmd{skip}}
\newcommand{\nstop}{\textbf{stop}}
\def \wait {\textbf{wait}}
\newcommand{\pstop}{\epsilon}
\newcommand{\evofun}[2]{{#1}(\dot{#2},#2)=0}
\newcommand{\evolution}[3]{\langle {#1}(\dot{#2},#2)=0 \& #3\rangle}
\newcommand{\evolutionn}[2]{\langle #1 \& #2\rangle}
\newcommand{\imitate}[4]{\langle {#1}(\dot{#2},#2)=0; #3 \& #4\rangle}
\newcommand{\pwait}{\textrm{wait}}
\newcommand{\exempt}[4]{#1 \unrhd \talloblong_{#2} (#3 \longrightarrow #4)}
\newcommand{\exemptS}[3]{#1 \unrhd (#2 \rightarrow #3)}
\newcommand{\external}[3]{\talloblong_{#1} (#2 \longrightarrow #3)}
\newcommand{\internal}[3]{\sqcup_{#1} (#2 \rightarrow #3)}
\newcommand{\lr}{\talloblong}
\newcommand{\evo}[3]{\langle \dot{#1}=#2\& #3\rangle}
\newcommand{\s}{\mathbf{s}}
\newcommand{\te}[1]{\texttt{#1}}
\newcommand{\ifelse}[3]{\texttt{if}\ #1 \ \texttt{then}\ #2 \ \texttt{else}\ #3}

 \newcommand{\aadlsstoc}{\mathit{AADL\!\oplus\!S/S2C}}

\newcommand{\added}[1]{\textcolor{blue}{#1}}
\newcommand{\deleted}[1]{\textcolor{red}{\st{#1}}}
\newcommand{\modify}[2]{\deleted{#1}\added{#2}}
\newcommand{\xdeleted}[1]{\textcolor{red}{\xcancel{#1}}}
\newcommand{\bdeleted}[1]{\textcolor{red}{\bcancel{#1}}}


\begin{abstract}
We introduce Mars 2.0 for modeling, analysis, verification and code generation of Cyber-Physical Systems. Mars 2.0 integrates Mars 1.0 with several important extensions and improvements, allowing the design of cyber-physical systems using the combination of AADL and Simulink/Stateflow ($\aadlss$), which provide a unified graphical framework for modeling the functionality, physicality and architecture of the system to be developed. For a safety-critical system, 
formal analysis and verification of its combined $\aadlss$ model can be conducted via the following steps. First, the toolchain automatically translates $\aadlss$ models into Hybrid CSP (HCSP), an extension of CSP for formally modeling hybrid systems. Second, the HCSP processes can be simulated using the HCSP simulator, and to complement incomplete simulation, they can be verified using the Hybrid Hoare Logic prover in Isabelle/HOL, as well as the more automated \textsf{HHLPy} prover. Finally, implementations in SystemC or C can be automatically generated from the verified HCSP processes. The transformation from $\aadlss$ to HCSP, and the one from HCSP to SystemC or C, are both guaranteed to be correct with formal proofs. This approach allows model-driven design of safety-critical cyber-physical systems based on graphical and formal  models and proven-correct translation procedures.  We demonstrate the use of the toolchain on several benchmarks of varying complexity, including several industrial-sized examples. 
\end{abstract}



\received{20 February 2007}
\received[revised]{12 March 2009}
\received[accepted]{5 June 2009}

\maketitle

\section{Introduction}

Cyber-physical systems (CPSs) seamlessly integrate computational and physical capabilities, that expand the capabilities of 
physical world through communication, computation and control. CPSs are omnipresent in our daily life  from aerospace, transportation, automotive, to health care, \emph{etc}. All these systems are expected to achieve desired behaviors, especially for safety-critical ones, for which any failure may cause severe catastrophes. 
Model-based design (MBD), which aims to design efficient and reliable systems, has become a popular development approach in CPS industry. It controls complexity of systems by appropriate abstraction/refinement and composition/decomposition, and meanwhile, it provides techniques based on (formal) modeling, analysis and verification in order to detect and correct errors in the early stage of design.
In the MBD approach, a system is usually modeled at different levels of abstraction for different purposes, e.g. graphical
models for engineering design and requirement analysis, formal models for formal verification, 
and executable code for implementation. However, these models at different levels must be consistent with each other to obtain a final correct implementation of the system. 

The MBD of CPSs faces the following challenge: a CPS tightly combines heterogeneous components related to physicality (including hardware platform and physical processes),  software and architecture, and their combinations.
To design a CPS at system level, it is desirable to take modeling, analysis, verification and implementation of these different aspects into account uniformly. Unfortunately, most of the existing MBD approaches do not support this. For example,  AADL (Architecture Analysis \& Design Language)~\cite{aadlAS5506C} supports the description of embedded system hardware and software architectures, relevant discrete behaviors, and their compositions based on connections. But it could not handle the continuous behavior of physical plants to be monitored and controlled by computers. On the other hand, Matlab's Simulink/Stateflow~\cite{slusing,sfusing} is an industrial model-based design environment that supports both discrete and continuous data flows, and flexible event-driven control based on hierarchical state machines. But it is not well-suited to describe system architectures and hardware platforms.
To address this challenge, recent work~\cite{XuWZJTZ22} investigates the combination of AADL and Simulink/Stateflow called $\aadlss$.  
$\aadlss$ provides a unified graphical framework for designing CPSs, which inherits the advantages of AADL and S/S, and avoids their disadvantages.

\begin{figure}[h!]
	\centering
	\includegraphics[scale=0.45]{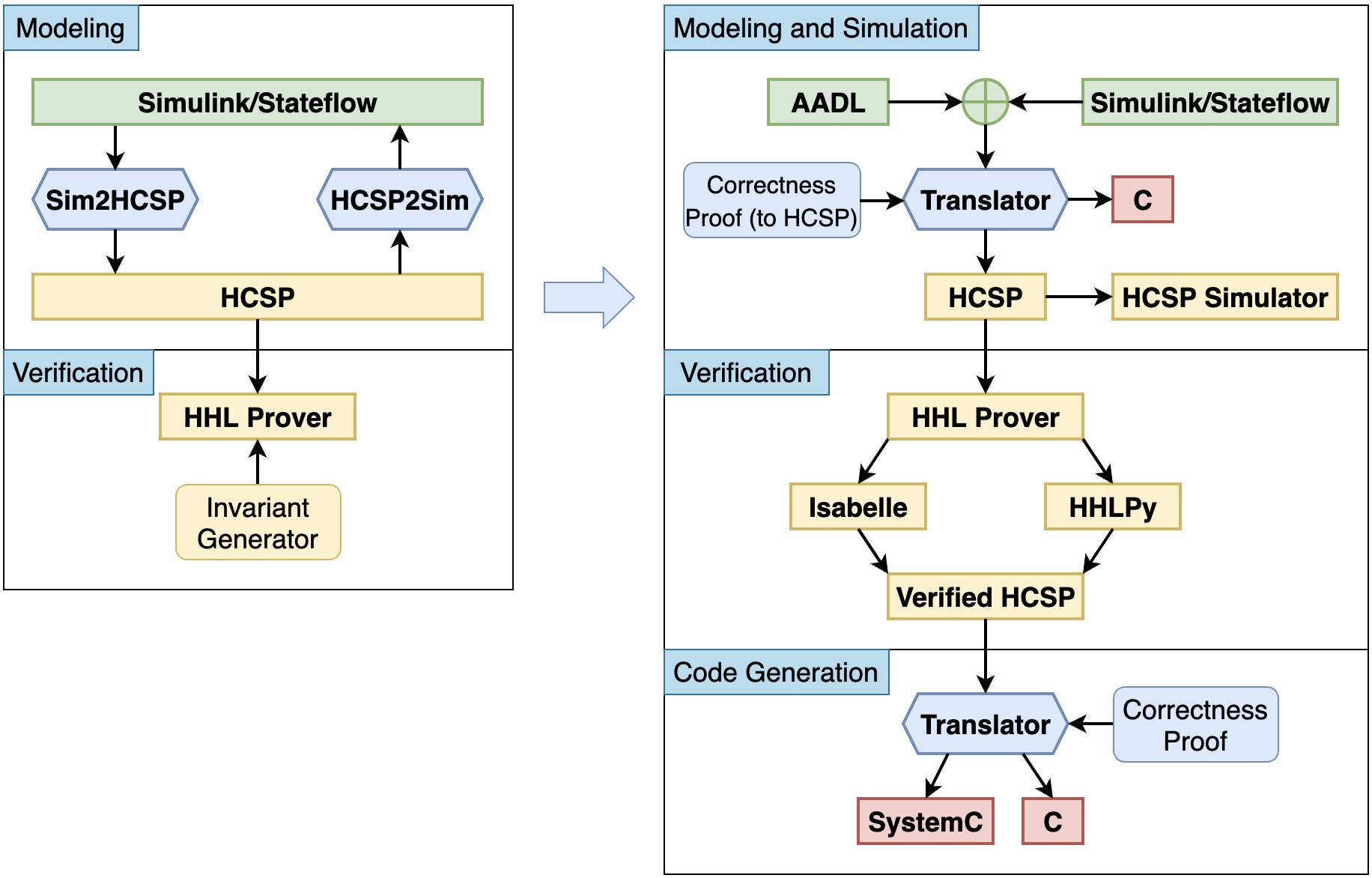}
 \caption{Mars 1.0 (left, \cite{DBLP:books/sp/17/ChenHTWYZZZ17}) and 2.0 (right)}
	\label{fig:mars}
\end{figure}

In addition to combination of AADL and Simulink/Stateflow, other aspects of the MBD approach, such as formal semantics of Simulink/Stateflow graphical models, formal verification~\cite{LiuLQZZZZ10,WangZZ15}, and code generation from formal models~\cite{DBLP:journals/tosem/YanJWWZ20} have also been investigated. Part of these works are implemented and combined into a toolchain called MARS~\cite{DBLP:books/sp/17/ChenHTWYZZZ17}, for \textbf{M}odelling, \textbf{A}nalysis
and Ve\textbf{R}ification of Hybrid \textbf{S}ystems. As shown in Fig.~\ref{fig:mars} left, MARS supports graphical modeling by Simulink/Stateflow and   automatic transformation from graphical models to HCSP formal models for verification. In particular, HCSP, which is an abbreviation for hybrid communicating sequential processes~\cite{Zhou95,He94}, is an extension of CSP with ordinary differential equations (ODEs) for modeling hybrid systems. 

Since the publication of~\cite{DBLP:books/sp/17/ChenHTWYZZZ17}, several extensions and improvements are added into the toolchain. Most significantly, it incorporates the combination of AADL and Simulink/Stateflow ($\aadlss$) in~\cite{XuWZJTZ22}, and the co-simulation of combined $\aadlss$ models by translation to C code and defining their integration~\cite{DBLP:conf/utp/ZhanLWTXZ19}. Other additions include a simulation tool for HCSP described in~\cite{DBLP:conf/utp/ZhanLWTXZ19}, and translation procedures from Simulink and Stateflow models into HCSP programs whose results are easier to understand and verify, as well as supporting more features (the procedure for Stateflow is described in~\cite{GuoZXWS22}, and the one for Simulink will be described in this paper). Finally, formal verification tools are significantly improved~\cite{hhlpy}, and a new code generation process to C is added.

Presentations of the above works are scattered in various papers (some of which still unpublished). Moreover, there is not yet a comprehensive overview of how the entire toolchain can be applied to facilitate the MBD approach. Hence, we present in this paper an updated version of the toolchain, called Mars 2.0. We give an overview of the new components since~\cite{DBLP:books/sp/17/ChenHTWYZZZ17}, as well as how the toolchain as a whole is used in practice. The architecture of the toolchain is shown on the right of Fig.~\ref{fig:mars}, where the green boxes (AADL and Simulink/Stateflow) indicate graphical models that serve as input to the toolchain; the yellow boxes indicate formal models (HCSP), and their simulation and verification tools; and the red boxes indicate generated code. The architecture of the previous version of this tool is shown on the left of Fig.~~\ref{fig:mars}, which is mainly composed of three parts: translators \textsf{Sim2HCSP} and \textsf{HCSP2Sim}, and an HHL prover. \textsf{HCSP2Sim} is used to justify the correctness of \textsf{Sim2HCSP}, and HHL prover verifies HCSP formal models. In Mars 2.0, the correctness of the translation procedures is proved formally, thus \textsf{HCSP2Sim} is no longer necessary.

In summary, we present in this paper a unified account of the Mars 2.0 toolchain, as well as show some examples of its application on industrial-sized case studies. Moreover, descriptions of the following aspects of the toolchain have not appeared in other papers.
\begin{itemize}
\item We present a new algorithm for transforming Simulink models to sequential HCSP (the original algorithm in~\cite{DBLP:books/sp/17/ChenHTWYZZZ17,XuWZJTZ22} results in parallel processes). This gives HCSP programs that are easier to verify.

\item We present some implementation details of code generation to C, in particular its support for extensions to HCSP. This includes preprocessing to handle channel parameters, and type inference for typing a well-formed HCSP program before it is transformed to C.

\item We describe how the toolchain can be used to support the MBD approach, including a realistically-scaled example on the whole design of the Automatic Cruise Control system, the simulation and verification of several complex case studies.
\end{itemize}


After reviewing related works in this section, the paper is organized as follows: Sect.~\ref{sec:Modeling} presents modeling frameworks  based on $\aadlss$ and HCSP respectively, and their simulation methods. Sect.~\ref{sec:tohcsp} presents the translation from $\aadlss$ to HCSP. Sect.~\ref{sec:hhlprover} presents the HHL prover for verifying HCSP models, including implementations in Isabelle/HOL and in Python. The code generation to C is presented in Sect.~\ref{sec:toccode}. We present experiments of using Mars 2.0 on several benchmarks in Sect.~\ref{sec:experiments}, and finally conclude in Sect.~\ref{sec:conclusion}.

\subsection{Related Work}
Except for the aforementioned closely-related work, 
there exist a huge amount of work on the modeling, analysis, verification and code generation of CPSs. We briefly summarize them from two directions, which consider CPSs from a complete or a separate view respectively. 

Several unified frameworks have been proposed for designing CPSs. The Metropolis design framework~\cite{metropolis,DavarePlatformbasedDesign} is a platform-based design environment for heterogeneous systems, that provides simulation, verification, and code synthesis by transforming all models to a unified meta-model language. However, it lacks support for physical plant modeling. 
Ptolemy~\cite{Ptolemaeus:14:SystemDesign} aims to design heterogeneous systems that combine different models of computation in terms of actors and provides modeling and simulation techniques for the combined models. Functional Mock-up Interface (FMI 3.0)~\cite{Junghanns21} is an industrial 
standard maintained by the Modelica Association that  enables
the exchange and co-simulation of dynamic component models. It couples different simulation tools at system level by coordinating and synchronizing their respective executions.  However, Ptolemy supports very limited facilities to model continuous
behaviors~\cite{Cremona17}, and furthermore, both Ptolemy and FMI are not designed for hardware architecture modeling and analysis.


Other lines of work consider limited perspectives of CPSs. UML, SysML~\cite{sysml} and MARTE~\cite{marte} are traditional model based design environments for designing discrete  systems, without support for physical plants. There are some works aiming for modeling and verifying continuous and hybrid behaviors, but without considering architectures. 
Z\'elus~\cite{HSCC2013} extends the synchronous language Lustre~\cite{lustre} with ODEs and zero-crossing events for designing and implementing hybrid systems. It supports analysis of hybrid models by type systems and   semantics~\cite{HSCC2013,JCSS2012}, and compilation that generates code for simulation and for embedded targets by source-to-source transformation~\cite{EMSOFT2011,CC2015}. Differential dynamic logic is developed for reasoning about behaviors of hybrid dynamic models~\cite{Platzer18}, and based on which the KeYmaera X prover ~\cite{DBLP:conf/cade/FultonMQVP15} is implemented for safety analysis of dynamic systems. Their later work transformed verified high-level models of CPSs in differential dynamic logic to executable controllers, with the safety properties preserved~\cite{DBLP:conf/pldi/BohrerTMMP18,DBLP:conf/iccps/GarciaMP19}. 
Though the correctness issue is addressed, only the discrete part of the model is considered. 

There are a number of commercial tools supporting code generation from dynamic and hybrid systems, such as Simulink~\cite{slusing}, AADL~\cite{aadlAS5506C,OSATE}, SCADE~\cite{dormoy2008scade}, Rational Rose~\cite{RATIONAL_ROSE}, and TargetLink~\cite{TargetLink} etc.
OSATE~\cite{OSATE} is the AADL tool environment, which provides modeling and analysis of real-time systems in AADL and furthermore supports the automated code generation from AADL models including runtime behavior and scheduling  to  C code.
However, they all validate  the process of the code generation by simulation, without formal guarantee for correctness.  
SCADE~\cite{dormoy2008scade} supports automatic code generation from their models and provides formal guarantee for correctness. It has been used for designing safety-critical embedded systems, but it is founded on the synchronous data-flow language Lustre~\cite{lustre}, which does not address continuous plants. Moreover, all of the above work provide no support for architecture modeling and analysis. 

\section{Modeling and Simulation}
\label{sec:Modeling}

This section reviews the existing work on modeling and simulation: graphical modeling based on $\aadlss$ and the co-simulation by transforming $\aadlss$ to C, formal modeling based on HCSP and the simulation of HCSP. 
\subsection{Graphical Modeling and Simulation with $\aadlss$}
\label{AADL2C}

Using $\aadlss$, a cyber-physical system is modeled with the following three layers:
\begin{description}
    \item[Architecture layer] The system architecture and its hardware platform are described by AADL components that define the structure, type and characteristics of composed hardware and software components.
    \item[Software layer] The software behavior can be modeled either through AADL behavioral annexes or S/S diagrams.
    \item[Physical layer] The physics of the cyber-physical system and its interaction with the hardware/software platform are modeled by S/S diagrams.
\end{description}

The simulation of $\aadlss$ models is performed by executing the C code generated from the graphical models. The C code generation is divided into the following three parts.

\paragraph{Translating AADL to C}

For each \textsf{thread} in the AADL part, a corresponding \texttt{Thread} object containing its component properties (such as dispatch protocol, priority, deadline, period, and minimum/maximum execution times) will be created. Besides the properties described in AADL, \texttt{Thread} also includes other properties like thread state.

\paragraph{Translating S/S to C}

Matlab provides an automatic code generation tool to translate S/S diagrams into C code that can simulate the model step-by-step. To apply the code generation tool, we need to set some configuration parameters, such as the step size, the ODE solver, format of the generated code, etc. The C code generated from an S/S diagram by the tool can be roughly divided into three functions: \texttt{initialization} (input), \texttt{computation} (execute for one step), and \texttt{finalization} (output). Thus, the behavior of a Thread object \texttt{thread} can be defined by the following function pointers:
\[
\begin{array}{l}
\texttt{thread->initialize = initialization;}\\
\texttt{thread->compute = computation;}\\
\texttt{thread->finalize = finalization;}\\
\end{array}
\]
where \texttt{initialization}, \texttt{computation}, and \texttt{finalization} are the functions in the C file generated by Matlab's code generator.

\paragraph{Simulation}

The C code of an $\aadlss$ model is combined together through a function implementing the thread scheduling protocol such as the Highest Priority First (HPF) protocol. The communication between components is implemented by shared variables in the context of C code. The step size of the Matlab simulation is set to agree with that of AADL simulation. The simulation result can then be visualized using Python’s plotting library, serving as a visual check that properties of the model are satisfied for the given initial state.

\begin{figure}
    \centering
    \includegraphics[scale=0.25]{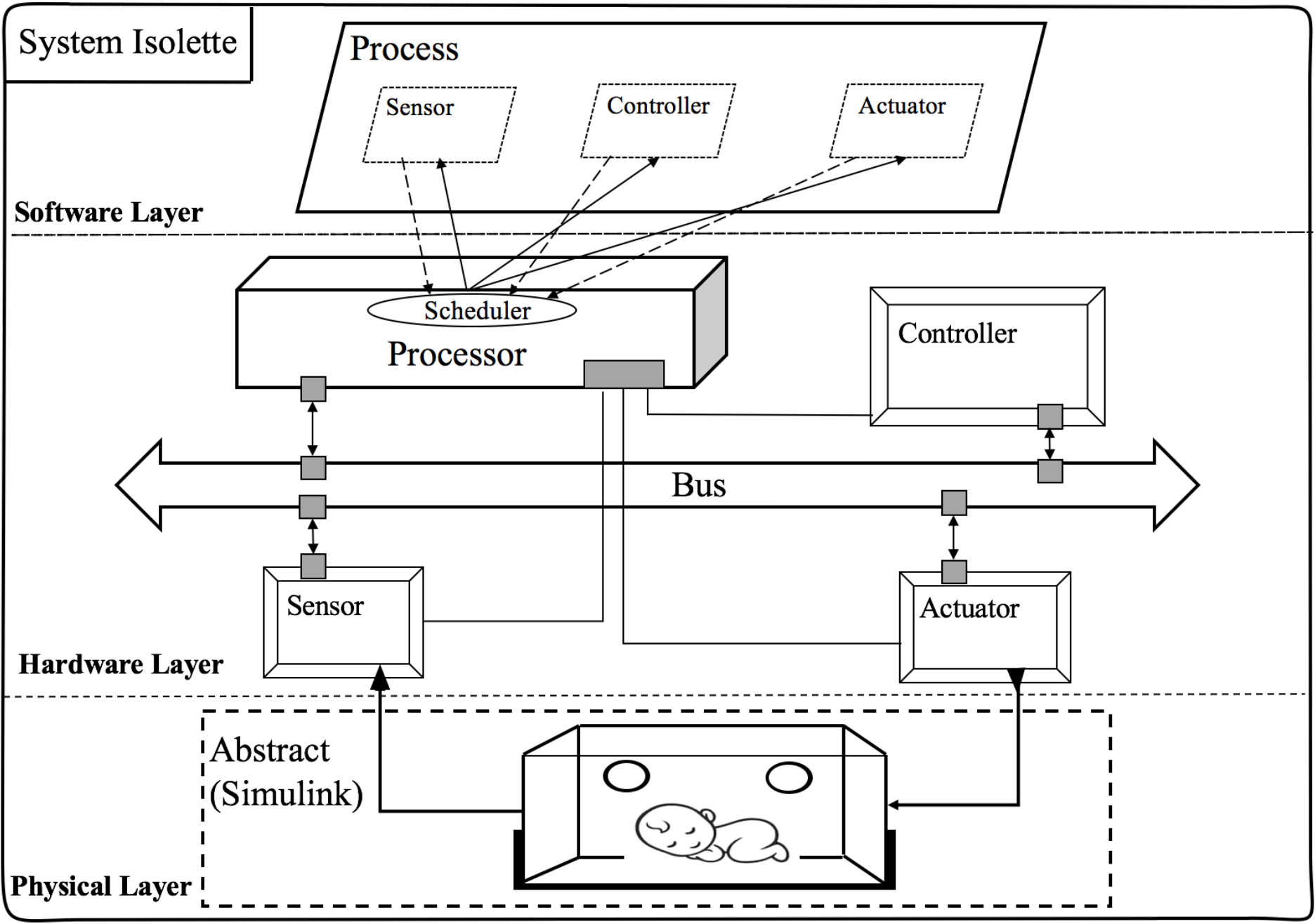}
    \caption{Graphical model of Isolette system (from~\cite{DBLP:conf/utp/ZhanLWTXZ19})}
    \label{fig:Isolette}
\end{figure}

Fig.~\ref{fig:Isolette} shows the $\aadlss$ model of an Isolette system which controls the temperature in an infant incubator. The system consists of one physical component describing the temperature evolution of the incubator, and this component is modeled by a Simulink diagram. The architecture part of the system is modeled using AADL. It consists of a central \textsf{Processor} with three component devices: \textsf{Sensor}, \textsf{Controller}, and \textsf{Actuator}. At the software layer, it consists of three threads with different priorities, which are scheduled according to the scheduling policy specified by the \textsf{Processor}. The simulation of this combined model can be found in~\cite{DBLP:conf/utp/ZhanLWTXZ19}.

\subsection{Formal Modeling and Simulation with HCSP}
\label{sec:hcsp}

Hybrid CSP (HCSP) is a formal language for describing hybrid systems which extends CSP by introducing differential equations for modeling continuous evolution, and interrupts for modeling the interaction between continuous evolution and communications. For full definition of HCSP, readers are referred to~\cite{He94,zhan2017formal}. Compared to the standard HCSP syntax, we make use of an extended language including datatypes such as strings and lists, operations on these datatypes, arrays of channels, and module definitions. 
We also allow definition of functions and procedures in a module. This makes translation of $\aadlss$ easier, but poses further challenges to simulation and code generation.

A simulator for the extended HCSP with a graphical
user interface has been implemented. The backend of the simulator is implemented in Python. In particular, solving of ODEs is done using Python's \texttt{scipy} package
(function \texttt{solve\_ivp}), which is also able to accurately
calculate the time at which the boundary of the domain is reached
using a root-finding algorithm. Finally, the simulator is linked to a
web interface which is able to show the HCSP process in pretty-printed
form, the steps of execution, and a plot of the variables in the
process against time. This allows us to not only view the result of
running an HCSP process, but also find out what went wrong if the
process does not execute as expected.

As shown in Sect.~\ref{sec:tohcsp}, Mars 2.0 also supports the simulation of $\aadlss$ models by translating $\aadlss$ to HCSP first and then doing the simulation on the resulted HCSP models. Compared to the simulation by translating $\aadlss$ to C, the second approach guarantees the correctness of translation formally and thus is more reliable.

\section{From $\aadlss$ to HCSP}
\label{sec:tohcsp}



In this section, we first review translation procedures from $\aadlss$ that are introduced in existing work~\cite{XuWZJTZ22}, where the translation procedures for Simulink and Stateflow are described in~\cite{ZouZWFQ13,ZouZWF15}. Then, we introduce the new algorithms that translate Simulink and Stateflow diagrams to sequential HCSP processes.

\subsection{Translation of $\aadlss$}
\label{sec:aadl_trans}

The translation of combined $\aadlss$ models to HCSP is introduced in~\cite{XuWZJTZ22}. We give a brief overview here.
In addition to the AADL model and associated Simulink/Stateflow models, a configuration file is required, supplying additional information such as the correspondence between channel names in the AADL model and input/output variable names in the HCSP or Simulink/Stateflow models.

\paragraph{Translating threads}

Each thread is translated into two HCSP processes. The \emph{dispatch} process is used to dispatch the thread, according to its dispatch protocol. Currently the \emph{periodic} and \emph{aperiodic} protocols are supported. The \emph{execution} process models execution of the thread. It represents a state machine with 4 states: \textsf{wait} (waiting to be dispatched), \textsf{ready} (dispatched but not running), \textsf{running} (currently running) and \textsf{await} (waiting for some resource, such as bus access). The execution behavior of a thread is specified using either HCSP code directly, or using Simulink/Stateflow diagrams. In the latter case, the procedures given in~\cite{ZouZWFQ13,ZouZWF15} are used to translate them into HCSP code, which is then spliced into the state machine model. In the latest version, proposed in this paper, we adopt the new translation algorithms for Simlulink and Stateflow, which will be introduced in Sect.~\ref{sec:simulink_trans} and Sect.~\ref{sec:stateflow_trans}.

\paragraph{Translating the scheduler}

The scheduler manages a group of threads and determines which thread to run at any time according to the scheduling protocol, which can be one of FIFO (first-in-first-out), RMS (rate-monotonic), DMS (deadline-monotonic) and HPF (highest priority first). For example, in the HPF case, each thread has a priority, and the scheduler maintains a pool of ready threads, at each time choosing the thread with the highest priority to run.

\paragraph{Translating devices and abstract components}

Devices and abstract components (for modeling physical processes) are expressed using HCSP code directly, or using Simulink/Stateflow diagrams. As with threads, procedures in Sect.~\ref{sec:simulink_trans} and Sect.~\ref{sec:stateflow_trans} can also be used to translate them into HCSP code, which is itself the translation of the component. 

\paragraph{Translating buses and connections}

Finally, we discuss translation of connections between threads. When two threads communicate directly without buses (e.g. when they are bound to the same processor), the connection is translated to a buffer modeled by a (stationary) ODE waiting for input and output communications. This models direct (asynchronous) communication between these threads.

When the communication is bound to one or more buses, additional code is required on the thread side to request and release bus access. Moreover, bus delay is modeled by inserting the corresponding delay in the implementation of the bus thread.

\subsection{Translation of Simulink}
\label{sec:simulink_trans}

The work~\cite{ZouZWFQ13} first introduced a translation algorithm from Simulink to HCSP, together with proof of correctness. In that work, a Simulink diagram is separated into discrete and continuous sub-diagrams, which are translated into separate HCSP processes. The entire Simulink diagram is then formed by parallel composition of these HCSP processes, with variables shared by the sub-diagrams transmitted by communication in HCSP. Communication is also used to model triggered subsystems. This method of translation has the advantage of putting plant and control in the Simulink model into separate processes. However, the extra communication complicates analysis and verification of the translate models, in particular those involving theorem proving.

The work~\cite{JLAMP2023} mentioned a different translation algorithm from Simulink to HCSP, that results in a sequential HCSP program. The focus of~\cite{JLAMP2023} is to define a denotational semantics for Simulink diagrams, and how this semantics can be used to prove the correctness of the translation from Simulink to HCSP using UTP theory, hence the translation itself is only described briefly and at a high level. In this paper, we give a systematic description of this translation algorithm. This algorithm differs from that in~\cite{ZouZWFQ13} mainly by removing the use of communication between discrete and continuous part of the diagram, as well as for triggered subsystems. This results in sequential programs that are easier to verify by theorem proving methods. Other improvements include giving a full treatment of discrete blocks with states, e.g. delay blocks, as well as computation blocks within the continuous part of the diagram. The Simulink blocks supported by our tool are summarized in Table~\ref{simBlocks}.

\begin{table}[h]
\caption{The Simulink blocks supported by Mars 2.0}
\label{simBlocks}
\begin{tabular}{|c|c|l|}
\hline
Category & Blocks & \multicolumn{1}{c|}{Description} \\ \hline
\multirow{5}{*}{Sources} & Constant & Outputs a constant signal. \\ \cline{2-3} 
 & Clock & Outputs the current simulation time. \\ \cline{2-3} 
 & Sine Wave & Outputs a sine wave. \\ \cline{2-3} 
 & Signal Builder & \begin{tabular}[c]{@{}l@{}}Creates and generates interchangeable groups of signals whose\\ waveforms are piecewise linear.\end{tabular} \\ \cline{2-3} 
 & \begin{tabular}[c]{@{}c@{}}Discrete Pulse\\ Generator\end{tabular} & Generates square wave pulses at regular intervals. \\ \hline
\multirow{2}{*}{Continuous} & Integrator & Continuous-time integration of the input signal. \\ \cline{2-3} 
 & Transfer Function & \begin{tabular}[c]{@{}l@{}}Models a linear system by a transfer function of the Laplace-\\ domain.\end{tabular} \\ \hline
\multirow{2}{*}{Discontinuities} & Hit Crossing & \begin{tabular}[c]{@{}l@{}}Detects when the input signal reaches the crossing offset value\\ in the direction specified by the crossing direction.\end{tabular} \\ \cline{2-3} 
 & Saturation & Limits input signal to the upper and lower saturation values. \\ \hline
\multirow{2}{*}{Discrete} & \begin{tabular}[c]{@{}c@{}}Discrete PID\\ Controller\end{tabular} & Implements the discrete-time PID control algorithm. \\ \cline{2-3} 
 & Unit Delay & Samples and holds with one sample period delay. \\ \hline
\multirow{2}{*}{\begin{tabular}[c]{@{}c@{}}Logical and Bit\\ Operations\end{tabular}} & Logical Operators & Logical operators including AND, OR, NOT, and so on. \\ \cline{2-3} 
 & Relational Operators & \begin{tabular}[c]{@{}l@{}}Applies the selected relational operator (such as $\leq$ and $>$)\\ to the inputs and outputs the result.\end{tabular} \\ \hline
\multirow{8}{*}{Math Operations} & Add & Adds or subtracts inputs. \\ \cline{2-3} 
 & Bais & Adds bias to input. \\ \cline{2-3} 
 & Gain & Multiplies the input by a constant value. \\ \cline{2-3} 
 & MinMax & Outputs min or max of input. \\ \cline{2-3} 
 & Abs & Outputs the absolute value of input. \\ \cline{2-3} 
 & Product & Multiplies or divides inputs. \\ \cline{2-3} 
 & Sqrt & Outputs the square root of input. \\ \cline{2-3} 
 & Square & Outputs the square of input. \\ \hline
\multirow{4}{*}{Signal Routing} & \begin{tabular}[c]{@{}c@{}}Data Store\\ Memory\end{tabular} & \begin{tabular}[c]{@{}l@{}}Defines and initializes a named shared data store,  which is a \\ memory region usable by Data Store Read and Data Store Write\\ blocks that specify the same data store name.\end{tabular} \\ \cline{2-3} 
 & Mux & Multiplexes scalar or vector signals. \\ \cline{2-3} 
 & Selector & Selects or reorders specified elements of a multidimensional input. \\ \cline{2-3} 
 & Switch & \begin{tabular}[c]{@{}l@{}}Passes through the first input or the third input signal based on the\\ value of the second input.\end{tabular} \\ \hline
\multirow{3}{*}{Subsystems} & Normal Subsystems & Contains a subset of blocks within a model or system. \\ \cline{2-3} 
 & Enabled Subsystems & \begin{tabular}[c]{@{}l@{}}If the value of the control signal is greater than zero,\\ the subsystem executes.\end{tabular} \\ \cline{2-3} 
 & Triggered Subsystems & \begin{tabular}[c]{@{}l@{}}The subsystem executes once the value of the control signal\\ crosses zero (falling, raising, or either).\end{tabular} \\ \hline
Sinks & Scope & Displays input signals with respect to simulation time. \\ \hline
\end{tabular}
\end{table}

\subsubsection{Causality graph}

We first review some important concepts about Simulink models, in particular the causality graph. A Simulink model consists of blocks connected together by wires. Each block may be classified as \emph{discrete} or \emph{continuous}. Discrete blocks update periodically according to its \emph{sample time}, while continuous blocks are updated continuously. Examples of discrete block include the gain block, which periodically compute its output as a certain multiple of its input, and the delay block, which delays its (discrete) input by one sample time. 

Any discrete block can be described mathematically using an internal state, together with \emph{update} ($f$) and \emph{output} ($g$) functions. Consider a discrete block with sample time $\mathsf{st}>0$ and internal states $\mathbf{s}$. Let $\mathbf{x}$ and $\mathbf{y}$ be the input and output signals of the block, respectively, then for each $k\in\mathbb{N}$, we have
\[
\begin{array}{rcll}
\mathbf{s}(k\cdot\mathsf{st}) &=& f\left(\mathbf{x}(k\cdot\mathsf{st}),\mathbf{s}((k-1)\cdot\mathsf{st})\right) & (\text{update})\\
\mathbf{y}(k\cdot\mathsf{st}) &=& g\left(\mathbf{x}(k\cdot\mathsf{st}),\mathbf{s}((k-1)\cdot\mathsf{st})\right) & (\text{output})
\end{array}
\]
For example, the gain block has no internal state or update function, and its output function is given by:
\[ 
\begin{array}{rcl}
y(k\cdot \mathsf{st}) &=& a \cdot x(k\cdot \mathsf{st})
\end{array}
\]
where $y$ is the output signal, $x$ is the input signal, $a$ is the gain ratio, and $\mathsf{st}$ is the sample time. The delay block has a single state variable $s$, with update and output functions as follows:
\[
\begin{array}{rcll}
  s(k\cdot \mathsf{st}) &=& x(k\cdot \mathsf{st}) & \text{(update)}\\
  y(k\cdot \mathsf{st}) &=& s((k-1)\cdot \mathsf{st}) & \text{(output)}
\end{array}
\]

The \emph{causality graph} of the discrete part of the Simulink diagram is defined as follows. Each node of the graph corresponds to a wire. Each block connects each output wire to the set of input wires that the output function depends on. For example, the gain block connects its output to its input, but the delay block introduces no edges to the graph (as its output depends only on the internal state). We assume the Simulink diagrams under consideration are well-formed, in the sense that its causality graph is acyclic~\cite{JLAMP2023}.

Examples of continuous blocks include the continuous version of the gain block, which always maintains its output as a certain multiple of its input, and the integrator block, which specifies its output as the integral of its input. Mathematically, the gain block can be defined as:
\[
\begin{array}{cl}
y(t) = a\cdot x(t) & \text{for all }t\ge0
\end{array}
\]
while the integrator block can be defined using an ODE:
\[ \dot{y}(t) = x(t) \]
There are other continuous blocks that result in more complex ODEs. For example, the transfer function block used in the powertrain control benchmark (Sect.~\ref{sec:powertrain}) with transfer function $\frac{1}{0.1s+1}$ is given mathematically by the following ODE: $\dot{y}(t) = -10y(t) + 10x(t)$.
In general, a continuous block may also contain internal state variables, with some output defined as function of internal and input variables.

The causality graph of the continuous part of the Simulink diagram is defined as follows. Again, each node in the graph corresponds to a wire. Each block connects each output wire to the set of input wires it depends on through an equation, rather than through an ODE. For example, the continuous gain block connects its output to its input, while the integrator block and the above transfer function block introduces no edges. The entire Simulink diagram is well-formed if the combined causality graph of its continuous and discrete parts is acyclic.

\subsubsection{Subsystems}

Simulink allows hierarchical design of models using \emph{subsystems}. They come in three types: normal subsystems, enabled subsystems, and triggered subsystems. All subsystems consists of blocks (and perhaps other subsystems) that can be considered as a single component of the model. Normal subsystems are executed at all times. Enabled subsystems are executed when a certain condition holds. Triggered subsystems are executed on the rising, falling, or either events of a triggering signal. We consider enabled and triggered subsystems consisting of discrete blocks only. The paper~\cite{JLAMP2023} gave the semantics and definition of causality graphs for all three types of subsystems.

\subsubsection{Preprocessing}

Before the main translation step, several preprocessing steps are performed on the Simulink diagram. First, normal and enabled subsystems are unfolded, with enabling condition recorded onto individual blocks of enabled subsystems. Next, we determine the sample times of blocks that are not explicitly specified. This is done by forward and backward propagation of sample times, according to the method given in the Simulink manual~\cite{slusing}. This classifies each block as either discrete or continuous. The sample time of the entire Simulink diagram is also determined at this step, as the greatest common divisor (GCD) of the sample times of the discrete blocks. The last preprocessing step separates the Simulink diagram into discrete and continuous parts, which are first translated individually (to be described below), then put together to form the translation result of the entire Simulink diagram.

\subsubsection{Translating Discrete Diagrams}\label{translate_dis}

Each wire and internal state of discrete block corresponds to a variable in the translated HCSP program. To each discrete block $\mathsf{B}$, we associate two HCSP code fragments $\mathsf{B}.O$ and $\mathsf{B}.U$ for output and update, respectively. For example, the gain block has:
\[
\begin{array}{ccll}
\mathsf{B}.O &\widehat{=}& y:=a\cdot x & (\text{output})\\
\mathsf{B}.U &\widehat{=}& \pskip & (\text{update})
\end{array}
\]
while the delay block has:
\[
\begin{array}{ccll}
\mathsf{B}.O &\widehat{=}& y:=s & (\text{output})\\
\mathsf{B}.U &\widehat{=}& s:=x & (\text{update})
\end{array}
\]

Translation of the discrete part of the diagram begins by a topological sort of the causality graph, giving a linear order on the wires. By identifying each block with its output wire, this gives a linear order on the blocks as well. Suppose the discrete blocks are $\mathsf{B}_1,\dots,\mathsf{B}_n$ according to this order, then the translation to HCSP first performs all output operations of blocks in order, then performs all update operations. That is:
\[ 
\begin{array}{rcl}
\mathsf{Discrete} &\widehat{=}&
\mathsf{B}_1.O;\cdots;\mathsf{B}_n.O;\mathsf{B}_1.U;\cdots;\mathsf{B}_n.U
\end{array}
\]

For blocks with sample time different from the sample time of the entire diagram, a condition need to be added to take this into account. Likewise, the enabling condition need to be added for blocks within enabled subsystems.

Some initialization steps are also needed for the discrete part of the diagram. In particular, each block with internal state require initialization of its state. Triggered subsystems and Stateflow charts (to be discussed later) may also require initialization. We collect initialization of discrete (and later continuous) part of the diagram into the code fragment \textsf{Init}.

\begin{figure}
    \centering
    \includegraphics[scale=0.45]{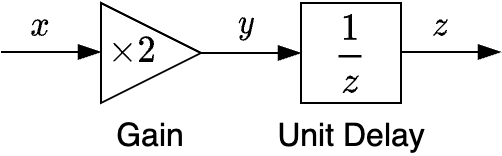}
    \caption{Example of discrete diagram}
    \label{fig:discrete_example}
\end{figure}

We give a concrete example to illustrate the translation procedure. Consider the diagram in Fig.~\ref{fig:discrete_example}. Suppose the topological order chosen for the causality graph is $z,x,y$, corresponding to the order \textsf{Unit Delay}, \textsf{Gain} on the blocks. Let $s$ be the internal variable of the unit delay block. Then the result of translation is:
\[
\begin{array}{rcl}
\mathsf{Discrete} &\widehat{=}& z := s; y := 2x; s := y
\end{array}
\]
and the initialization is: $\mathsf{Init_d}~\widehat{=}~s := 0$.

\subsubsection{Translating Continuous Diagrams}

Each continuous block not involving ODEs can be specified by a function $g$ from its input $\mathbf{x}$ to its output $\mathbf{y}$. Therefore, such blocks can be represented by a variable substitution $\mathbf{y}\mapsto g(\mathbf{x})$. Let
\[
\begin{array}{rcl}
G &\widehat{=}& \{\mathbf{y}_1\mapsto g_1(\mathbf{x}_1), \cdots, \mathbf{y}_n\mapsto g_n(\mathbf{x}_n)\}
\end{array}
\]
be the set of variable substitutions. Since the causality graph of the Simulink diagram is acyclic, there are no cycles in these variable substitutions, hence they form a well-defined substitution $\Gamma$ by composition.

\begin{figure}[h!]
    \centering
    \includegraphics[scale=0.45]{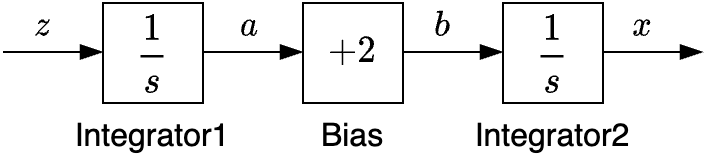}
    \caption{Example of continuous diagram}
    \label{fig:continuous_example}
\end{figure}

Next, the translation of the continuous diagram consists of collecting together ODEs of the blocks, then performing the variable substitution $\Gamma$. This is represented as the ODE $\dot{\mathbf{y}}=\Gamma(\mathbf{x})$. The initialization $\mathsf{Init_c}$ gives the initial value of ODEs.
We illustrate this with an example. Consider the diagram in Fig.~\ref{fig:continuous_example}.
The two integrator blocks yield the ODE $\dot{a}=z, \dot{x}=b$. The bias block yields the substitution $b\mapsto a+2$. Hence, the result of translation is the ODE $\dot{a}=z, \dot{x}=a+2$. The initialization is: $\mathsf{Init_c}~\widehat{=}~a:=0; x:=0$.

\subsubsection{Translating Triggered Subsystems}

Triggered subsystems can be considered as a stateful block with two internal variables \textit{pre} and \textit{triggered}, to record respectively the last value of the trigger line and whether the subsystem was triggered at the last round. Initially, \textit{pre} is set to 0 and \textit{triggered} is set to $\mathbf{false}$.

Let $\mathbf{x}$ and $\mathbf{y}$ be the input and the output of the triggered subsystem, and $\mathit{cur}$ be the value of the trigger line. The trigger conditions $\mathit{trig}(\mathit{pre}, \mathit{cur})$ is defined as follows, depending on whether the trigger type is rising, falling, or either:
\[
\begin{array}{ll}
(\mathit{pre}<0\wedge\mathit{cur}\geq0) \vee (\mathit{pre}\leq0\wedge\mathit{cur}>0) & (\text{rising})\\
(\mathit{pre}>0\wedge\mathit{cur}\leq0) \vee (\mathit{pre}\geq0\wedge\mathit{cur}<0) & (\text{falling})\\
(\mathit{pre}<0\wedge\mathit{cur}\geq0) \vee (\mathit{pre}>0\wedge\mathit{cur}\leq0) & (\text{either})\\
\vee(\mathit{pre}=0\wedge\mathit{cur}>0)\vee(\mathit{pre}=0\wedge\mathit{cur}<0) &
\end{array}
\]
Let $\mathbf{y}\coloneqq g(\mathbf{x})$ represent the computation performed by the (discrete) blocks in the triggered subsystem. At each round, this computation is performed if the subsystem was not triggered in the previous round ($\neg\mathit{triggered}$) and the trigger condition holds ($\mathit{trig}(\mathit{pre},\mathit{cur})$). Then, we update the values of $\mathit{pre}$ and $\mathit{triggered}$ accordingly. In summary, the \textsf{output} and \textsf{update} processes of a discrete triggered subsystem \textsf{DTrig} can be defined as follows:
\[
\begin{array}{rcl}
\textsf{DTrig}.O &\widehat{=}& \neg\mathit{triggered}\wedge\mathit{trig}(\mathit{pre}, \mathit{cur})\to\mathbf{y}\coloneqq g(\mathbf{x})\\
\textsf{DTrig}.U &\widehat{=}& (\mathit{triggered}, \mathit{pre})\coloneqq(\textsf{TrigCond}, \mathit{cur})
\end{array}
\]

\subsubsection{Translating Simulink Diagrams}

Translation of the entire Simulink diagram is formed by combining translation of its discrete and continuous parts. First, initialization of the discrete and continuous parts are performed. Then, a loop is entered where every round consists of the computation of the discrete diagram, followed by evolution of the continuous diagram for $d$ time units, where $d$ is the sample time of the entire diagram. That is: 
\[
\begin{array}{rcl}
\mathsf{Diagram}&\widehat{=}&
\mathsf{Init_d}; \mathsf{Init_c};
\left(\!\!\!
\begin{array}{l}
\textsf{Discrete};t\coloneqq0;\\
\langle\dot{\mathbf{y}}=\Gamma(\mathbf{x}), \dot{t}=1\&t<\mathit{period}\rangle
\end{array}
\!\!\!\right)^\ast
\end{array}
\]
For example, for the combination of the diagrams in Fig.~\ref{fig:discrete_example} and Fig.~\ref{fig:continuous_example}, the result of translation is (with sample time 1):
\[
s:=0; a:=0; x:=0; (z:=s; y:=2x; s:=y; t:=0; \ode{\dot{a}=z, \dot{x}=a+2, \dot{t}=1}{t<1})^*
\]

\subsection{Translation of Stateflow}
\label{sec:stateflow_trans}

At a basic level, Stateflow diagrams consists of \emph{states} and \emph{transitions} between states. However, there are many additional features in Stateflow to provide various conveniences in modeling. They include hierarchical states, inter-level transitions, junctions, events and temporal events, messages, and so on. Moreover, Stateflow permits states to contain evolution by ODEs, making it a convenient tool for modeling hybrid systems, in particular switched systems. The existing work~\cite{GuoZXWS22} describes in detail a translation procedure from Stateflow diagrams to HCSP, covering all of the above features. In this section, we give a brief overview of the main ideas, and an example of translation of ODEs in Stateflow, a feature that is used in the case study in Section~\ref{sec:cruise-controller}.

The translation is organized into two stages: a syntax-directed translation stage, and a code-optimization stage. In the syntax-directed translation stage, each element in the Stateflow chart is translated into a single procedure according to the Stateflow semantics. For example, for each state, three procedures are created for the entry, during, and exit process of the state. The entry procedure calls the \textsf{en}-action of the state, then calls the entry procedures of its child states (in addition to updating some book-keeping variables). The during procedure tries the outgoing transitions, \textsf{du}-action of the state, inner transitions, and during procedures of active child states, in that order. The exit procedure of a state first performs the \textsf{ex}-action of the state, then calls the exit procedure of its child states. An event stack \texttt{EL} is used to keep track of current events. Raising an event is translated to first pushing the event onto the stack, call the execution of either the full chart (in the case of broadcast events) or a particular state (for directed events), and finally popping the event from the stack.

The above syntax-directed translation procedure is easy to understand and prove correct. However, it results in HCSP processes that consist of many procedure definitions, as well as other extraneous code. The second part of translation is a code-optimization stage. In this part, various optimization techniques common to compilers are performed. They include inlining of procedures, peephole optimization, constant propagation, and dead code elimination. The Stateflow translation is evaluated on 112 hand-designed examples, validating consistency of the semantics by comparing simulation results in Simulink and of the translated HCSP program.

\subsubsection{Translation of continuous evolution}

Stateflow permits specifying ODEs within a state. This can be combined with other features of Stateflow, in particular hierarchical states, resulting in potentially complex semantics. The translation from Stateflow to HCSP collects together ODEs within each state across the hierarchy, and determines which ODE to execute according to the current state. The boundary of each ODE correspond to conditions on outgoing transitions from the corresponding state.

\begin{figure}
    \centering
    \includegraphics[scale=0.35]{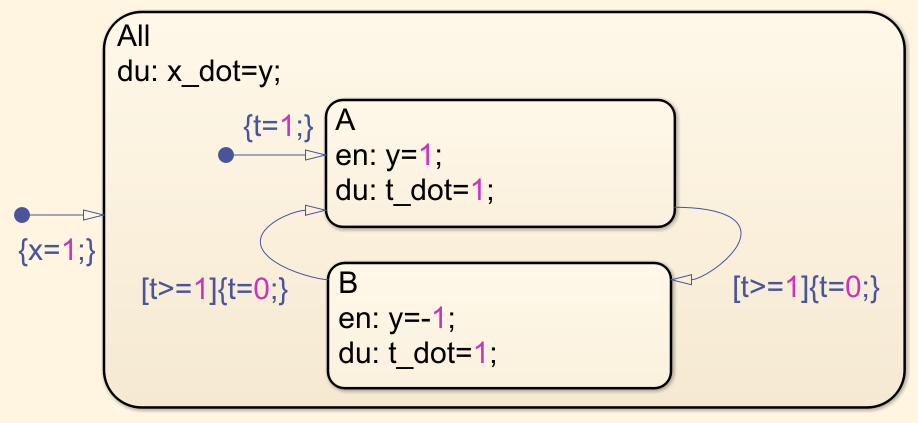}
    \caption{Sawtooth example}
    \label{fig:sawtooth}
\end{figure}

\paragraph{Example} Consider the Stateflow diagram shown in Fig.~\ref{fig:sawtooth} (reproduced from~\cite{hhlpy}). This diagram combines hierarchical states with specification of ODE within a state. The outer state specifies that variable $x$ evolves according to $\dot{x}=y$. The inner states specify that variable $t$ evolves according to $\dot{t}=1$ and modifies the value of $y$ on entry. Hence, the behavior of this diagram has $y$ alternately taking values $1$ and $-1$, while $x$ evolves in a sawtooth pattern. The translated HCSP contains the ODE evolution $\ode{\dot{x}=y,\dot{t}=1}{t<1}$ (where $t<1$ is the negation of the condition on the outgoing transitions), together with code specifying updates to the variable $y$ on transition between states $A$ and $B$.

\section{Verification}
\label{sec:hhlprover}

In this section, we describe verification of HCSP processes by theorem proving. Verification using HHL Prover is based on hybrid Hoare logic, with two modes having been implemented, one in the proof assistant Isabelle, and one for sequential HCSP processes only but provides more automation.

\subsection{Verification in Isabelle}

HHL Prover in Isabelle is implemented based on the hybrid Hoare logic (HHL) for deductive verification of HCSP processes. HHL is first introduced and implemented in~\cite{LiuLQZZZZ10,WangZZ15}, with assertions defined using Duration Calculus (DC)~\cite{DBLP:series/eatcs/ChaochenH04} and first-order logic. However,   DC-based proof systems for HCSP are too complicated to be used in practice. 

The HHL in Mars 2.0 is based on a recent improved proof system, mainly with the following differences: First, the semantics for HCSP processes is defined using traces, which record the ordered sequence of events consisting of communications and continuous evolution. Especially, each continuous evolution event records the readiness for potential communications. Traces can be combined in parallel with each other while guaranteeing that synchronized communications occur at the earliest possible time. Second, assertions in the HHL are defined as purely first-order predicates on traces. Based on the new assertions, for each construct in HCSP, a corresponding Hoare rule is defined in the weakest-precondition form. In particular, rules for combining assertions for parallel programs are defined based on the trace synchronizations, allowing us to conclude the specification of parallel processes from those  of individual, sequential processes.

The prover in Isabelle has been applied to the case studies of lunar lander~\cite{ZhaoYZGZC14}, Mars lander~\cite{ZhanGXJW0LCYZ21}, and Chinese Train Control System (CTCS-3)~\cite{ZouZWFQ13}. More recently, we used the prover based on the new HHL to verify the correctness of the composition of one scheduler with two tasks for AADL thread scheduling.

\subsection{Verification using HHLPy}
\label{sec:hhlpy}

\textsf{HHLPy} is a prover with a friendly graphical user interface for verifying properties of sequential HCSP processes~\cite{hhlpy}. The sequential part of HCSP processes contains discrete computation and continuous evolution statements, but not communication, interrupts or parallel processes. Properties specified in first-order logic are annotated as pre- and postconditions. Together with other annotations such as invariants and proof rules for ODEs, \textsf{HHLPy} can generate verification conditions automatically. These verification conditions are sent to the SMT solver Z3~\cite{MouraB08} or Wolfram Engine. If all verification conditions are proved, the HCSP process is shown to satisfy the desired properties.

The Hoare logic rules in \textsf{HHLPy} include regular rules for reasoning about discrete programs, as well as rules for reasoning about ODEs such as differential weakening, differential invariant, differential cut, and differential ghost, which are adapted from differential dynamic logic~\cite{Platzer18,PlatzerT20} to the case of HCSP processes. The main issue that need to be addressed is the difference in semantics between HCSP and differential dynamic logic concerning the boundary of ODE evolution. For this, a new form of judgment, called \emph{invariant triple} $\inv{P}{\dot{\boldsymbol{x}}=\boldsymbol{e}}{Q}$ is defined, stating that $Q$ is an invariant of the ODE $\dot{\boldsymbol{x}}=\boldsymbol{e}$ under domain $P$. The rules of reasoning about ODEs are then defined in terms of both invariant triples and regular Hoare triples.

\begin{figure}
    \centering
    \includegraphics[scale=0.1]{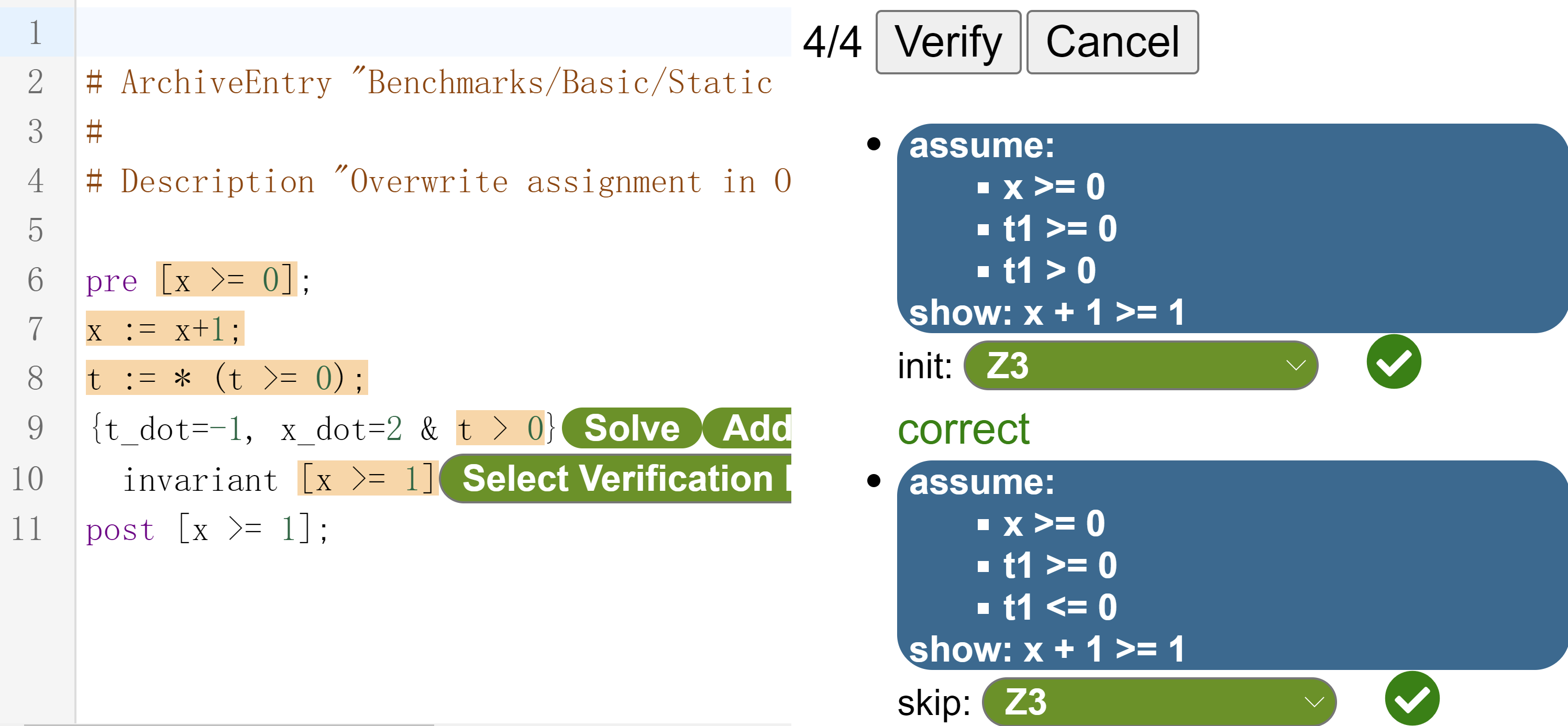}
    \caption{Screenshot of user interface. The left panel is the editor area, where user can edit HCSP processes and annotations. The right panel shows verification conditions with their labels, solvers selected and results.}
    \label{fig:hhlpy}
\end{figure}

\textsf{HHLPy} is implemented in Python, with user interface implemented in JavaScript. Proof rules and the verification condition generation algorithm are implemented in the Python-based core engine. Fig.~\ref{fig:hhlpy} shows a screenshot of the user interface. Specifically, the tool implements highlighting mechanisms to visualize different verification conditions. As shown in Fig.~\ref{fig:hhlpy}, relevant fragments for generating each verification condition are highlighted when hovering over it. In addition, the tool implements a labeling mechanism to distinguish between different verification conditions for proof reuse.

\section{C Code Generation}
\label{sec:toccode}

Automatic code generation forms the last part of the toolchain, which allows obtaining implementations of the system that are formally guaranteed to agree with the formal model. Code generation from HCSP to SystemC is introduced in~\cite{DBLP:journals/tosem/YanJWWZ20}, which gives a proof of correctness stated in terms of approximate bisimulation between the HCSP model and the generated SystemC code. In this section, we introduce code generation to C that is newly added in Mars 2.0. Compared to~\cite{DBLP:journals/tosem/YanJWWZ20}, we handle extensions to the HCSP language including different data types, module and channel parameters, and procedure definitions. Moreover, the synchronized communications of HCSP are implemented in C using the concurrency primitives in the POSIX pthreads library. In contrast to SystemC, which provides the same synchronized communication mechanisms as in HCSP, making their implementation more direct, the implementation of synchronized communication in C requires more care.

In other work currently under review, we presented the translation algorithm from HCSP to C, in particular how the synchronized communication in HCSP can be implemented using the pthreads library. We then prove a theoretical result stating the correctness of the implementation, in the sense of a bisimulation between the HCSP primitives and the translated concurrent C code under the interleaving model. 
The other part of the translation, discretization of ODEs, stays the same, and hence inherits the correctness result in terms of approximate bisimulation. The overall conclusion is that the original HCSP code, including evolution following ODEs and interrupts, is approximately bisimilar to the generated C code.

We now give an overview of the code generation procedure to C, focusing on the implementation aspects.

\subsection{Preprocessing}

The first step performs several preprocessing steps to remove aspects of the extended HCSP language that cannot be translated directly. This includes in particular module and channel parameters. The translation from $\aadlss$ to HCSP in Sect.~\ref{sec:aadl_trans} results in many processes that are instantiated from module templates. For example, the scheduler is parameterized by the processor ID, and the system may contain multiple schedulers instantiated from the same module with different processor IDs. In the first stage, we instantiate the modules into concrete HCSP processes.

Next, the translated HCSP code contains channels with parameters, which can both be matched in the external choice, and also be instantiated with variables. This is illustrated with the following (simplified) code fragment from the scheduler for the HPF protocol:

{\small
\begin{verbatim}
reqProcessor[0][_tid]?prior -->
    Pool_now := push(Pool_now, _tid);
    run_now := ...;
    run[0][run_now]!;
\end{verbatim}
}
\noindent Intuitively, it says that whenever the scheduler on processor 0 receives a request from thread \textit{tid}, indicating it is ready to run, the scheduler adds the thread to the pool of ready threads and recomputes the next thread to run based on priority, then sends the run signal to that thread via a communication. Both the matching on \textit{tid} and the variable parameter \textit{run\_now} pose problems for code generation, as the latter assumes static channels. Hence, when the set of all threads on processor 0 is known, the above code fragment is expanded to include one communication for each thread. For example, if there are two threads with ID ``t1'' and ``t2'', then the above code is expanded to:

{\small
\begin{verbatim}
reqProcessor[0]["t1"]?prior -->
    Pool_now := push(Pool_now, "t1"); run_now := ...;
    if (run_now == "t1") { run[0]["t1"]!; }
    else if (run_now == "t2") { run[0]["t2"]!; }
reqProcessor[0]["t2"]?prior -->
    Pool_now := push(Pool_now, "t2"); run_now := ...;
    if (run_now == "t1") { run[0]["t1"]!; }
    else if (run_now == "t2") { run[0]["t2"]!; }
\end{verbatim}
}
\noindent After this expansion, all communications are on channels with concrete parameters.

\subsection{Type inference}

The extended HCSP language contains variables with several different types, including numbers, strings, lists, and so on. The simulator described in Sect.~\ref{sec:hcsp} is dynamically typed, figuring out types of variables at runtime. However, the same cannot be done for code generation to the C language. Hence, before code generation, type inference is performed on the entire HCSP program. This includes identifying types of each variable in each thread, as well as types of values communicated on each channel. We make the restriction that each variable in each thread can hold values of at most one type, and each channel can communicate values of at most one type. Type inference may need to be performed in multiple rounds, as knowledge of types may be passed between threads through a channel. The type inference procedure flags an error if the above typing constraints are found to be violated, and also when it cannot conclude the type of some variable. In the latter case, extra initialization statements can be added for that variable to help the type inference procedure.

\subsection{From HCSP to C}

The main part of code generation translates each HCSP statement to the corresponding C code. For discretization of ODEs, we follow existing work~\cite{DBLP:journals/tosem/YanJWWZ20}: When generating the C code, the step size of discretization is determined with respect to the given precision, then the state of the continuous evolution at the next time point is computed by the Runge-Kutta method. Thus, the C code generated by the continuous evolution statement $\evolution{F}{\s}{B}$ is a loop structure that applies the Runge-Kutta method at each iteration, then waits for a step length of time before the next iteration, until the boundary condition $B$ becomes false. The C code generated by the continuous interrupt statement $\exempt{\evolution{F}{\s}{B}}{i\in I}{ch_i\triangleright}{P_i}$ is also a loop structure, where each iteration performs a discrete version of the interrupt, which waits for the availability of communications among $\{ch_i\}_{i\in I}$ as well as for a time limit. If some communication can occur, the loop breaks and continues to implement the corresponding $P_i$; otherwise, the state of continuous variables is updated and the loop is repeated, until $B$ becomes false. 

This stage results in C code that calls primitives corresponding to the implementation of time delay, input/output communications, and (discrete) interrupt.
We provide a set of functions that implement the above time delay and communication primitives using the \textsf{pthread} library.  
There are two variants of implementation which differ in how to deal with passage of time. In the real time variant, actual system delays are inserted into the C code, so the execution of C code proceeds at the same speed as the original HCSP model. In the virtual time variant, execution of C code proceeds as fast as possible, which can be used as a simulator and obtains simulation results quickly.

We now briefly explain the implementation of communication primitives, leaving the details (as well as correctness proof) to another paper. The implementation maintains several global data structures that are protected by a single lock. All functions below acquire the lock at the beginning and release it at the end  or when waiting for some event (time limit or communication) to occur.

\paragraph{Time delay}

To realize the progress of time in each thread and synchronize time between different threads, a local clock is introduced to represent the time in each thread, and a global clock is introduced to represent the current global progress of simulation. The implementation of $\mathsf{delay}(i,d)$ increases the local clock of thread $i$ by $d$, and then waits for the global clock to catch up.

\paragraph{Input and output}

Each channel is represented by a data structure with three fields: sender or receiver's token, channel content and channel direction. Each token defines whether the corresponding sender or receiver is available to participate in a communication. This takes the value of thread ID if some thread is ready to communicate, and $-1$ if otherwise.

Output $ch!e$ performs the following actions in sequence: set the token of $ch!$ to the thread number, put message $e$ into channel $ch$, read the token of $ch?$ to see if it is available. If yes, perform the communication immediately, wait for the input side to perform corresponding actions, and then set the token of $ch!$ back to $-1$; otherwise, the thread will wait for $ch?$ to become available and then perform the communication. The input call $ch?x$ performs symmetric actions, except when the communication occurs, it will read the value from channel $ch$ and assign it to variable $x$. In both cases, the thread releases the lock on the global data structures once to wait for the other side to perform corresponding actions, and the token in the channel reserves the communication so it cannot be interfered by other threads.

\paragraph{Interrupt}

Interrupt combines the actions of time delay and waiting for a list of communications $\{io_i\}$. It performs the following actions: first check and modify the tokens corresponding to each communication $\{io_i\}$. If any communication is ready, it is performed and the call returns with the index of communication, to allow the corresponding ensuing action $P_i$ to be performed. Otherwise it increments the local clock by the time delay of the interrupt, waits for all communications as well as for the global clock to catch up. If a communication occurs first, the local clock is set back to the value of global clock at the time, and the call returns the index of communication. If the global clock catches up first, the call returns to indicate no communication is performed within this step. In the case of communication, wait for the other side to perform corresponding actions as in the description of input and output above.

\paragraph{Main function}

The main function of the generated code first initializes the global variables, and then starts one thread for the code generated from each HCSP process. After all threads finished running, it releases resources for the global variables.

\section{Experiments}
\label{sec:experiments}

In this section, we describe several case studies of using the toolchain in practice. The toolchain, as well as all of the case studies, are available for download at \texttt{https://github.com/intgq/Mars-2.0}.

\subsection{Examples in Existing Work}

We first give a brief summary of existing case studies using the Mars toolchain.

In~\cite{ZouZWFQ13,ZouZWF15}, modeling and verification of the Chinese High-speed Train Control System (CTCS) are studied using pure Simulink/Stateflow environment for modeling and then transforming to HCSP for verification. The verification shows that the train can successfully complete the level transition from CTCS-2 to CTCS-3 and the mode transition from full supervision to calling on simultaneously in a combined scenario. 

In~\cite{ZhaoYZGZC14}, the modeling and verification of the slow descent phase of lunar lander are studied. First, a closed loop model consisting of the lander’s physical dynamics and the program controlling the lander is built; Second, the safety property is formally defined that if the initial conditions stay in a reasonable range, then the velocity during the landing must be maintained around a given value; Finally, the property is formally verified for the given model. In~\cite{ZhanGXJW0LCYZ21}, the authors apply the Mars toolchain to model the descent guidance control phase of the recently launched Tianwen I mars lander, and verify that it correctly controls the velocity of the lander.


\subsection{Simulation of Simulink/Stateflow Models}
\label{sec:powertrain}

\begin{figure}
    \centering
    \includegraphics[scale=0.65]{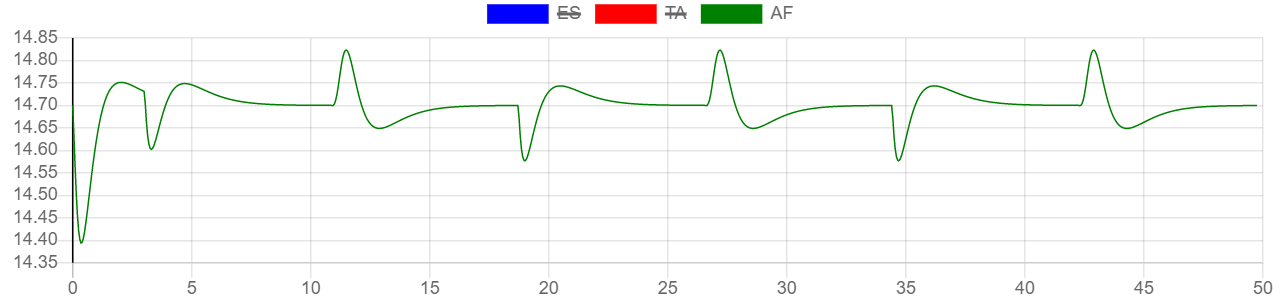}
    \caption{Simulation result of powertrain control in the HCSP simulator. The x-axis is time and the y-axis is AF (air-fuel ratio)}
    \label{fig:powertrain_result}
\end{figure}

We tested the procedure for translating Simulink/Stateflow models in Section~\ref{sec:simulink_trans} and~\ref{sec:stateflow_trans} on some more complex examples. The Powertrain control verification benchmark~\cite{JinDKUB14} consists of a series of models of powertrain control in automobile systems. The simpler model already uses many Simulink features, such as transfer functions (Transfer Fcn), Mux and Selectors (whose explaination can be found in Table~\ref{simBlocks}) for dealing with vectors of signals, as well as Matlab expressions to specify polynomials. The Yoyo control example is representative of models where a Stateflow diagram is placed in the middle of a Simulink model, used to provide control input. It also contains Simulink blocks such as hit-cross that generates an event whenever some signal crosses a given threshold. We translated both models into HCSP, and performed simulation and code generation. The results of simulating the HCSP code in the powertrain control model is shown in Fig.~\ref{fig:powertrain_result}. This agrees with the result of running the generated C program, as well as the simulation result within Simulink.

\subsection{Translation and Verification of Switched Systems}
\label{sec:cruise-controller}

In this part, we present a case study on modeling switched systems using Stateflow, followed by verification using \textsf{HHLPy}. The automatic cruise controller~\cite{Oehlerking11} is a switched system with six modes of operation and eleven transitions between them, which include PI controller, acceleration, two service brakes and two emergency brakes. Stability of the controller has been verified using KeYmaera X in~\cite{TanMP22}. Since \textsf{HHLPy} currently only supports verifying safety properties, we verify instead a rephrasing of the stability condition using safety properties.

\begin{figure}[h!]
    \centering
    \includegraphics[scale=0.6]{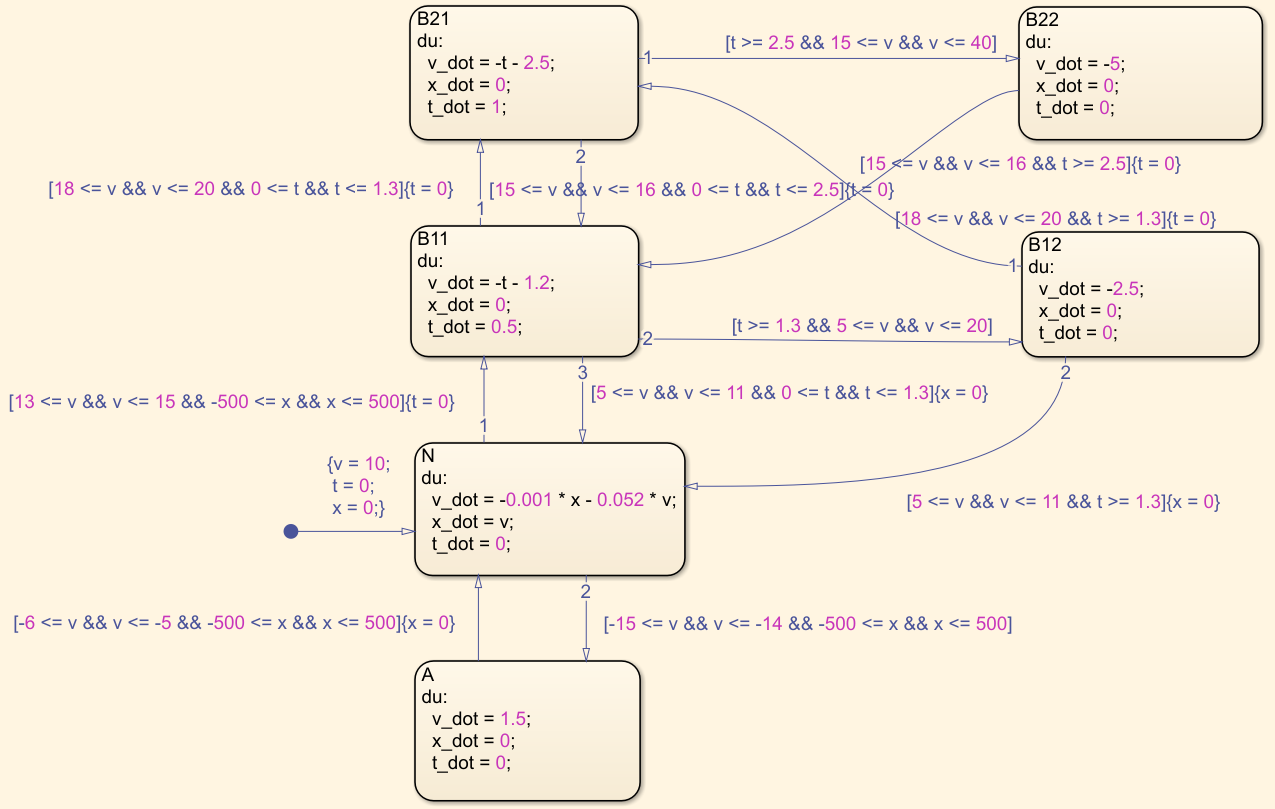}
    \caption{Stateflow Model of Automatic Cruise Control}
    \label{fig:ss_cruise}
\end{figure}

We modeled the system using Stateflow charts, shown in Fig.~\ref{fig:ss_cruise}. The Stateflow model is slightly different from the original model, as Stateflow models generally have deterministic behavior. Despite these modifications, the model captures the essential behaviors of the original model. The Stateflow model was translated into an HCSP process using the algorithm in Section~\ref{sec:stateflow_trans}.

We then verified the following safety property of the HCSP process: given that initial velocity satisfies $v_0^2 < \delta^2$ with $\delta$ sufficiently small ($\delta > 0 \wedge \delta \le 11 \wedge \delta < \epsilon - 0.1$) and initial mode is PI controller (state $N$), then velocity satisfies $v^2 < \epsilon^2$ throughout for all $\epsilon > 0.1$. This safety property is similar but weaker than stability (which impose conditions for all $\epsilon > 0$). The verification steps are similar to that in~\cite{TanMP22} for stability, concerning the PI controller only. Invariants for the system are that the system always stays in state $N$, together with a polynomial constraint based on the Lyapunov function of state $N$, which is verified by differential invariant rule. Due to the many states actually present in the system, a total of 1288 verification conditions are generated, all of which are checked correct using the Z3 solver.

\subsection{The Whole Design of Automatic Cruise Control System}


For the last case study, we adopt the realistically-scaled Automatic Cruise Control System (ACCS for short) from~\cite{XuWZJTZ22} to explain the whole procedures. The translation from $\aadlss$ to HCSP and the verification for the generated HCSP model were addressed in~\cite{XuWZJTZ22}. In this case, we add a new part: from HCSP to C (Sect.~\ref{case_hcsp_c}).

\begin{figure}
\centering
\includegraphics[scale=0.45]{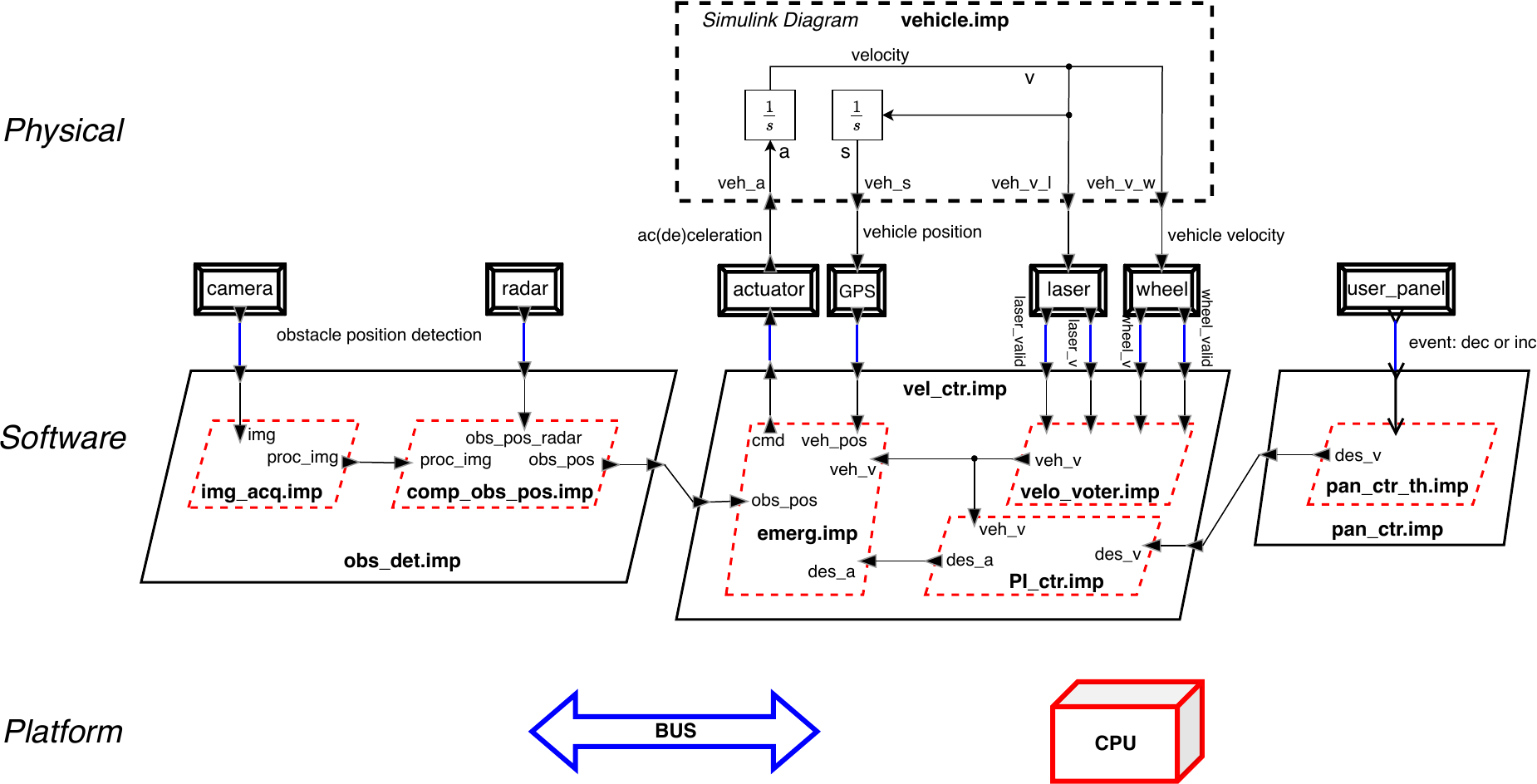}
\caption{Automatic Cruise Control System (borrowed from~\cite{XuWZJTZ22})}
\label{fig:CCS}
\end{figure}

The ACCS is modeled using $\aadlss$ in~\cite{XuWZJTZ22} and decomposes into three layers, as shown in Fig.~\ref{fig:CCS}. The physical layer is the physical vehicle described by a Simulink diagram. The software level defines control of the system. It contains three processes for obstacle detection, velocity control, and panel control, and each process is composed of several threads. These processes interact with the environment through devices. The platform layer consists of bus and processor. The connections in the software level can be bound to buses in the platform level.
All threads are bound to the single processor, with HPF scheduling policy. 

The execution of ACCS is as follows. A vehicle is placed at the starting point initially and the driver can accelerate (\texttt{inc}) and decelerate (\texttt{dec}) the vehicle by the \texttt{user\_panel}. Thread \texttt{pan\_ctr\_th} deals with the commands from the driver and then sends desired velocities (\texttt{des\_v}) to the discrete PI controller (thread \texttt{PI\_ctr}). Meanwhile, process \texttt{obs\_det} detects the obstacles ahead by a \texttt{camera} and a \texttt{radar} and provides the velocity controller process (\texttt{vel\_ctr}) with the real-time position of obstacle (\texttt{obs\_pos}). Thread \texttt{velo\_voter} in process \texttt{vel\_ctr} monitors the velocity of the vehicle using \texttt{laser} and another device located on one \texttt{wheel} of the vehicle and produces the real-time velocity of the vehicle (\texttt{veh\_v}) to the discrete PI controller \texttt{PI\_ctr} and the emergency control thread (\texttt{emerg}). Based on the real-time velocity of vehicle and the desired velocity received, \texttt{PI\_ctr} computes a desired acceleration (\texttt{des\_a}) which will be sent to \texttt{emerg}. Finally, \texttt{emerg} collects the real-time position (\texttt{veh\_pos} by \texttt{GPS}) and velocity of the vehicle, the desired velocity set by the driver, and the real-time position of the obstacle to work out a \texttt{command}, by some emergency control strategy, which will update the acceleration of the vehicle through \texttt{actuator}. The vehicle moves according to the new acceleration and the above procedure repeats. 

\subsubsection{Translation from $\aadlss$ to HCSP}

In~\cite{XuWZJTZ22}, the $\aadlss$ model of the ACCS of Fig.~\ref{fig:CCS} is stored in one JSON file, which was taken as an input by the translator to generate files of HCSP modules as outputs. In this work, however, the $\aadlss$ model of the ACCS is composed of three parts: (1) AADL files describing the three levels of architecture; (2) Simulink/Stateflow diagrams of XML format modeling the behaviours of threads, devices, and physical environment; and (3) a configuration file (introduced at the beginning of Sect.~\ref{sec:aadl_trans}) of JSON format. The translator takes these files as inputs and
generates files of HCSP modules of TXT format as outputs, where the translations for Simulink and Stateflow diagrams adopt the new algorithms introduced in Sect.~\ref{sec:simulink_trans} and~\ref{sec:stateflow_trans}. Readers could refer to~\cite{XuWZJTZ22} for more details of the translation procedures. 
Notice that the architecture of the ACCS could be changed (such as adding a bus) to obtain
variants of the model by modifying the AADL files directly.

\subsubsection{Verification for HCSP}

A generated HCSP model should be verified against some desired properties before being translated to executable code. Since the HCSP model of the ACCS of Fig~\ref{fig:CCS} is extremely complex, it was abstracted as a simple parallel composition of a plant and a controller in~\cite{XuWZJTZ22} and then the safety property (the vehicle will never collide onto the moving obstacle ahead and its velocity will never exceed the upper limit) could be verified using HHL in Isabelle/HOL.

\subsubsection{Translation from HCSP to C}
\label{case_hcsp_c}

The above obtained HCSP files can be fed into the translator introduced in Sect.~\ref{sec:toccode} to generate C code. 
This part is not included in the previous work of~\cite{XuWZJTZ22}.
The communication between processes is implemented using $\mathsf{pthreads}$ and the correctness of the code generation is proved based on approximate bisimulation, i.e., an HCSP model and the C code generared from it are in some approximate bisimulation relation.
The HCSP model is specified by a formal language and therefore verifiable. Thus, the C code generated from HCSP files is more reliable than from the graphical $\aadlss$ model directly (Sect.~\ref{AADL2C}).
The generated C code is of 3500--4000 lines, longer than the C code (about 2500 lines) generated by~\cite{DBLP:conf/utp/ZhanLWTXZ19} (Sect.~\ref{AADL2C}). One major reason is that the generated C code in this paper is approximately bisimilar to the original $\textsf{AADL}\oplus\textsf{S/S}$ model of ACCS, which means the detailed behaviours of ACCS are reflected in the C code and vice versa, while it is not the case for~\cite{DBLP:conf/utp/ZhanLWTXZ19} where no such bisimulation can be guaranteed. 

\subsubsection{Comparison}

We consider the following scenario. At the beginning, the driver pushes the \texttt{inc} button three times with time interval $0.5$s in between to set a desired speed to $3$m/s. After $30$s, the driver pushes the \texttt{dec} button twice in $0.5$s time intervals to decrease the desired speed. On the obstacle side, we assume that the obstacle appears at time $10$s and position $35$m, then moves ahead with velocity $2$m/s, before finally moving away at time $20$s and position $55$m. 

We adopt the parameters setting of Table 1 in~\cite{XuWZJTZ22} for the threads and devices in the ACCS of Fig.~\ref{fig:CCS}, except that the deadline of thread \texttt{pan\_ctr\_th} is shrunk to $50$ms in this paper.
We compare our results with the work~\cite{DBLP:conf/utp/ZhanLWTXZ19} (Sect.~\ref{AADL2C}) and the simulation of the HCSP model of ACCS~\cite{XuWZJTZ22}. It should be noted that bus latency is not considered in~\cite{DBLP:conf/utp/ZhanLWTXZ19}. Thus, to make it fair, we only consider the case with on bus latency, i.e., we ignore the \texttt{bus} component in Fig.~\ref{fig:CCS}. Examples involving bus latency can be found in~\cite{XuWZJTZ22}.

\begin{figure}[h]
\centering
\includegraphics[scale=0.46]{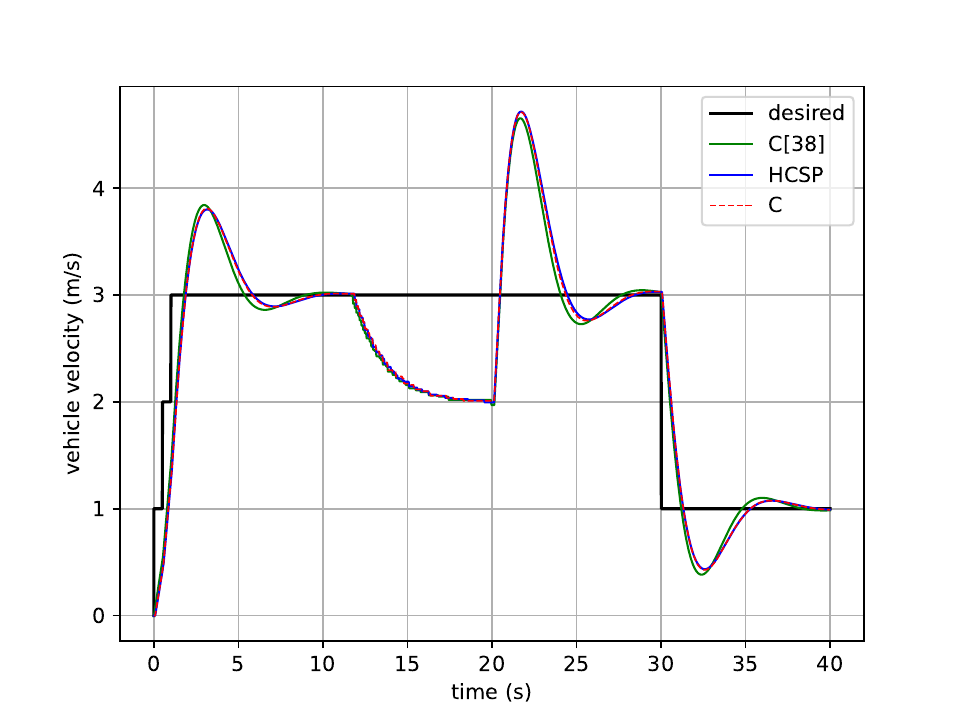}
\includegraphics[scale=0.46]{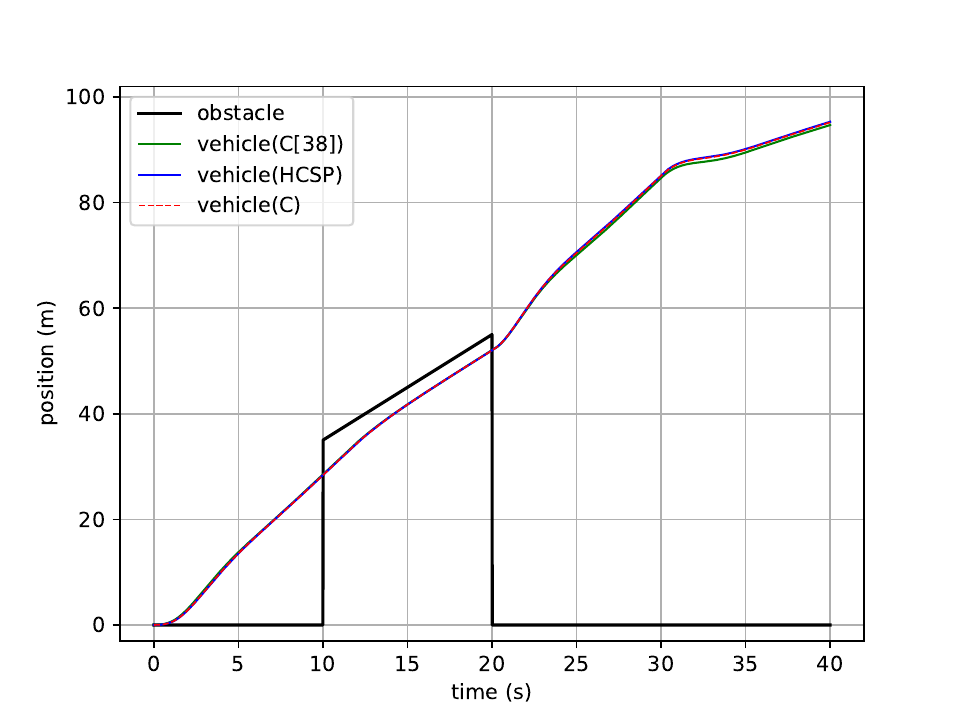}
\caption{Comparison of execution results}
\label{fig:veh_v_s}
\end{figure}

The left of Fig.~\ref{fig:veh_v_s} shows the evolutions of the vehicle speed, where the black line denotes the desired velocity set by the driver, and the red, green, and blue lines denote the results of our work, the work of~\cite{DBLP:conf/utp/ZhanLWTXZ19} (Sect.~\ref{AADL2C}), and the simulation of the HCSP model of ACCS~\cite{XuWZJTZ22}, respectively. We can see that the execution of the generated C code (red line) is almost the same with the simulation  of its HCSP model (blue line).

From all the three results, we can see that the vehicle accelerates to the desired speed ($3$m/s) in $10$s. The fluctuation during $[2\text{s},10\text{s}]$ reflects feature of PI controllers.
It then decelerates to avoid the collision onto the obstacle ahead. After the obstacle moves away (at $20$s), the vehicle accelerates again to the desired speed. At $30$s, the driver pushes the \texttt{dec} button to adjust the desired velocity to $1$m/s and we can see that the vehicle decelerates to $1$m/s in about $6$s under the PI controller. The positions of the vehicle and of the obstacle with respect to time are shown on the right of Fig.~\ref{fig:veh_v_s}.

\section{Conclusion}
\label{sec:conclusion}
This paper presents Mars 2.0, an integrated toolchain for formal design of CPSs, supporting to take  physicality, software and architecture into account uniformly. Mars 2.0 covers the whole design process of CPSs, from graphical modeling in $\aadlss$ at the beginning, to formal modeling in HCSP for simulation and  verification, and to final code implementation with correctness guarantee. 
It realizes the automatic transformation from $\aadlss$ to HCSP, and from HCSP to SystemC/C, both of which have formal correctness guarantees by proving the  semantic consistency between corresponding source and target models. Therefore, a more reliable CPS can be developed by using Mars 2.0 compared with existing tools. We also demonstrate its use on several case studies of varying complexity. 

As future work, we will try to integrate more functionalities into 
Mars 2.0, such as modeling and verification of hybrid systems containing delay or stochastic differential equations. We are also considering to implement a more friendly graphical user interface for the co-modeling of $\aadlss$, which currently only supports the textual descriptions of models. Finally, we intend to apply the toolchain to real-world case studies on a larger scale.

\bibliographystyle{ACM-Reference-Format}
\bibliography{reference}

\end{document}
\endinput